# Sparse Multipath Channel Estimation Using Compressive Sampling Matching Pursuit Algorithm

Guan Gui[1,2], Qun Wan[1], Wei Peng[2] and Fumiyuki Adachi[2]

1. *Dept. of Electric Engineering, University of electrical Science and Technology of China, Chengdu, 611731, China*
2. *Dept. of Electrical and Communication Engineering, Graduate School of Engineering, Tohoku University, Sendai, 980-8579, Japan*

*Correspondence author: gui@mobile.ecei.tohoku.ac.jp*

***Abstract*-Wideband wireless channel is a time dispersive channel and becomes strongly frequency-selective. However, in most cases, the channel is composed of a few dominant taps and a large part of taps is approximately zero or zero. To exploit the sparsity of multi-path channel (MPC), two methods have been proposed. They are, namely, greedy algorithm and convex program. Greedy algorithm is easy to be implemented but not stable; on the other hand, the convex program method is stable but difficult to be implemented as practical channel estimation problems. In this paper, we introduce a novel channel estimation strategy using compressive sampling matching pursuit (CoSaMP) algorithm which was proposed in [1]. This algorithm will combine the greedy algorithm with the convex program method. The effectiveness of the proposed algorithm will be confirmed through comparisons with the existing methods.**

## I. INTRODUCTION

Coherent detection in wideband mobile communication systems often requires accurate channel state information at a receiver. The study of channel estimation for the purposes of channel equalization has a long history. In many studies, densely distributed channel impulse response was often assumed. Under this assumption, it is necessary to use a long training sequence. In addition, the linear channel estimation methods, such as least square (LS) algorithm, always lead to bandwidth inefficiency. It is an interesting study to develop more bandwidth efficient method to acquire channel information.

Recently, the compressive sensing (CS) has been developed as a new technique. It is regarded as an efficient signal acquisition framework for signals characterized as sparse or compressible in time or frequency domain. One application of the CS technique is in channel estimation. If the channel impulse response follows sparse distribution, we can apply the CS technique. As a result, the training sequence length can be shortened compared with the linear estimation methods. Recent measurements show that the sparse or approximate sparse distribution assumption is reasonable [2, 3]. In other words, the wireless channels in real propagation environments are characterized as sparse or sparse clustered; these sparse or clustered channels are frequently termed as a sparse multi-path channel (SMPC). An example of SMPC impulse response channel is shown in Fig.1. Recently, the study on SMPC has drawn a lot of attentions and concerning results can be found in literature [4-6]. Correspondingly, sparse channel estimation technique has also received considerable interest for its advantages in high bit rate transmissions over multipath channel [7].

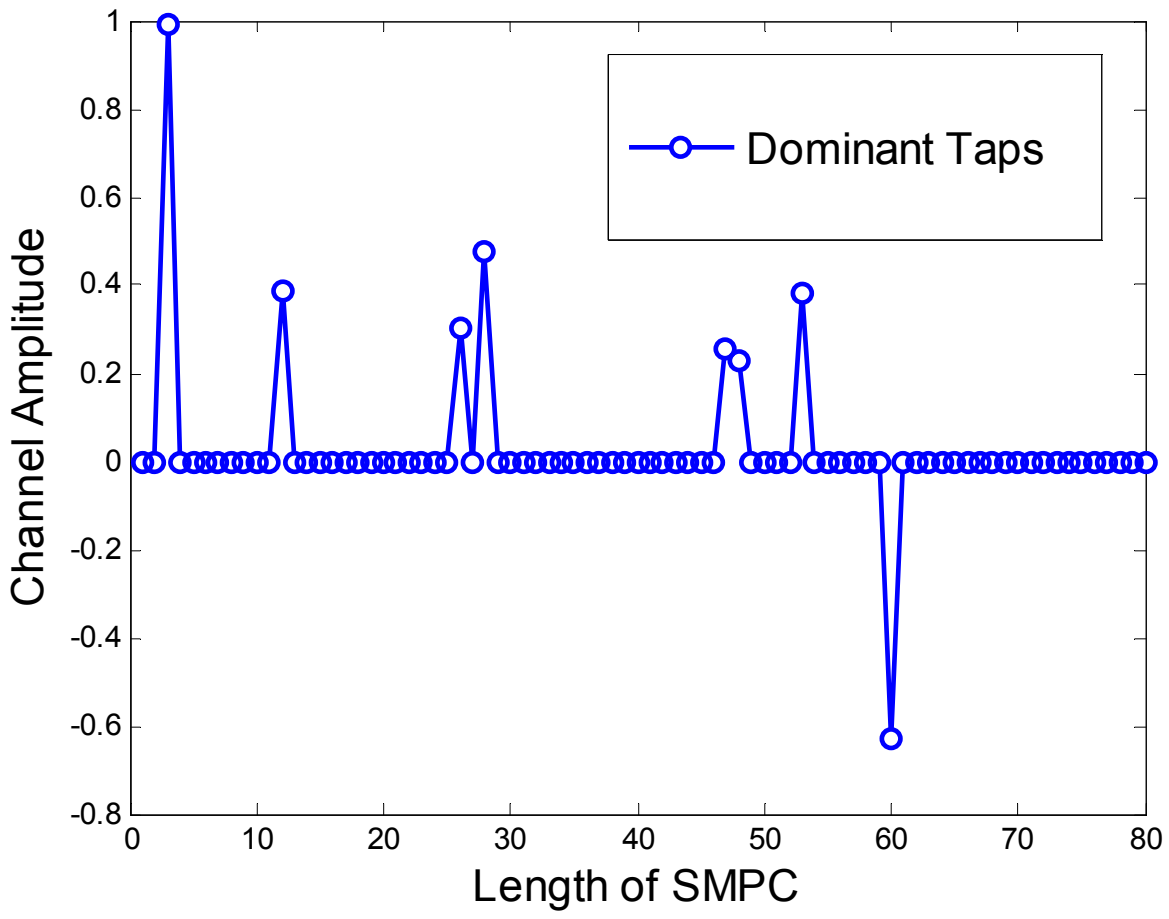

*Fig.1. An example of SMPC. Channel length is 80. The dominant taps distribute uniformly and the number of dominant taps is 8. It was necessary to state that we consider the real baseband channel model while neglect the imaginary part for simplification.*

Exploiting the sparse property of SMPC, orthogonal matching pursuit (OMP) algorithm [8, 9] and convex program algorithm [10] have been proposed. OMP algorithm is fast and easy to be implemented. However, the stability of OMP for sparse signal recovery has not been well understood yet. Donoho *et al* [11] suggested that OMP should be less stable than the convex program. Tropp and Gilbert [9] investigated the performance of OMP algorithm by the measurement. It is reported in [9] that if the channel matrices satisfy some properties, then OMP algorithm can recover the sparse signals with high probability. However, the conditions on OMP estimation algorithm are more restrictive than the restricted Isometry condition (RIC) [12]. Kunis and Rauhut [13] mathematically proved that the first iteration of OMP can identify the maximum dominant taps of channel with the training sequences. However, because of the unavoidable correlation between columns of the training sequence, the instability of the OMP algorithm easily leads to weak channel estimation.

Convex program method can resolve the instability of OMP algorithm. Convex program algorithm, such as Dantzig Selector (DS) [14], is based on linear programming. The main advantage of convex program method is its stability and high estimation accuracy. The convex problem method can work correctly as long as the RIC conditions are satisfied. However, this method is computationally complex and difficult to be implemented [15].

In this paper, we will introduce a novel SMPC estimation method using compressive sampling matching pursuit (CoSaMP) algorithm [1]. It has both the advantages of the greedy algorithm and the convex program. In other words, CoSaMP algorithm combines low computational complexity and robustness on practical channel estimation. The study in [1] focused on mathematical description on CoSaMP algorithm for sparse or approximate sparse signal recovery problem. And the perfect channel state information (CSI) was assumed while practical channel estimation was not considered. In this paper, we will use the CoSaMP algorithm to deal with the practical channel estimation problems.

The rest of the paper is organized as follows. Sparse multipath channel model is presented in Section II. Section III will describe the existing CoSaMP algorithm and propose a new SMPC estimation method by using the CoSaMP algorithm. In section IV, we will compare the performance of the proposed method with the existing methods by simulations. Finally, conclusions are drawn in Section V.

## II. SMPC Model

At first, the symbols used in this paper are described as following. The superscript $T$ stands for transposition. Bolded capital letters denote a matrix where bolded lowercase letters represent a vector. Notation $|\cdot|$ stands for the absolute value. Norm operator $\|\bullet\|_0$ denotes $\ell_0$ vector norm, i.e., the number of non-zero entries of the vector; $\|\bullet\|_1$ denotes $\ell_1$ vector norm, which is the sum of the absolute values of the vector entries. $\|\bullet\|_2$ denotes $\ell_2$ norm. $\tilde{\mathbf{h}}$ and $\mathbf{h}$ indicate estimate channel vector and actual channel vector, respectively.

We consider single-antenna wideband propagation systems, which are often equivalent to frequency-selective baseband channel model. An $N$-length training sequence $\mathbf{X}\in\mathbb{R}^{N*L}$ $(N<L)$ will be transmitted over a random stationary SMPC. The equivalent baseband transmitted $\mathbf{X}$ and received signals $\mathbf{y}=\left[y_0 \quad y_1 \quad \cdots \quad y_{N-1}\right]^T$ is given by

$$y[t]=\sum_{\tau} x[t-\tau]h[t,\tau]+z[t] \tag{1}$$

where $\tau$ is delay spread of multipath signal which is characterized by channel length $L$. Its matrix form can written as

$$\mathbf{y}=\mathbf{Xh}+\mathbf{z} \tag{2}$$

Where

$$\mathbf{X}=\begin{bmatrix} x_{00} & x_{01} & \cdots & x_{0(N-1)} & \cdots & x_{0(L-1)} \\ x_{10} & x_{11} & \cdots & x_{1(N-1)} & \cdots & x_{1(L-1)} \\ \vdots & \vdots & \ddots & & \ddots & \vdots \\ x_{(N-1)0} & x_{(N-1)1} & \cdots & x_{(N-1)(N-1)} & \cdots & x_{(N-1)(L-1)} \end{bmatrix}$$

denotes the training sequence and read its as row vector form $\mathbf{X}=\left[\mathbf{x}_0,\mathbf{x}_1,\ldots,\mathbf{x}_{N-1}\right]^T$ ; $\|\mathbf{h}\|_0=S$ represents the number of dominant channel taps of the SMPC. Suppose that there are $S$ dominant channel taps distributed randomly over the channel, according to the SMPC definition, $S\ll L$ ; $\mathbf{z}(t)=\left[z_0 \quad z_1 \quad \cdots \quad z_{N-1}\right]^T$ is the additive white Gaussian noise (AWGN) with zero mean and variance $\sigma^2$.

## III. CoSaMP Algorithm for SMPC Estimation

In this part, we will introduce some properties of compressed sensing (CS) theory as a basis for the CoSaMP channel estimation method. And then we will show how to apply the CoSaMP algorithm to sparse channel estimation.

### *A. Compressed Sensing*

Consider the baseband channel model of (1). If we want to guarantee accurate channel estimator, the training sequence $\mathbf{X}$ must satisfy two conditions.

*1) Restricted Isomery property (RIP) [12].*

The $S$-RIC of a $N\times L$ training sequence $\mathbf{X}$, denoted by $\delta_S$, is defined as the smallest value $\delta_S$ ($\delta_S\in(0,1)$) which can satisfy the inequality

$$(1-\delta_S)\|\mathbf{h}\|_2^2\le\|\mathbf{Xh}\|_2^2\le(1+\delta_S)\|\mathbf{h}\|_2^2 \tag{3}$$

for any SMPC vector $\mathbf{h}$. If (2) is satisfied, the training sequence $\mathbf{X}$ is said to satisfy RIP of order $S$ and accurate channel estimator can be obtained by using CS methods. From CS perspective, research on the RIP of the training sequence has two important purposes. First, RIP-based training sequence is a sufficient condition to robust probe sparse channel dominant taps. Furthermore, in the process of error performance analysis, RIC of training sequences play important role to improve lower bound.

*2) Lower bound of length of training sequence* $\mathbf{X}$

Due to the channel fading and noise, how to determine the length of training sequence $\mathbf{X}$ is important in terms of both spectrum efficiency and estimation robustness. Therefore, the length $N$ of $\mathbf{X}$ must satisfy [16]

$$N\ge C_1\cdot S\cdot(\log L)^4\,\mu_{\mathbf{X}}^2, \tag{4}$$

where $C_1$ is a constant and $\mu_{\mathbf{X}}=\sqrt{L}\max_{i,j}\left|\mathbf{X}_{i,j}\right|$ which is known as the maximum coherence between the $i^{th}$-column and $j^{th}$-column of $\mathbf{X}$

### *B. CoSaMP SMPC*

***Theorem1.*** *If N-length training sequence* $\mathbf{X}$ *satisfies the RIP and* $\delta_{2S}\le\sqrt{2}-1$ *[12], for any 2S-sparse channel vector* $\mathbf{h}$, *CoSaMP algorithm produces the channel estimator* $\tilde{\mathbf{h}}$ *that satisfies*

$$\|\mathbf{h}-\tilde{\mathbf{h}}\|_2\le C\cdot\max\left\{\varepsilon, 1/\sqrt{S}\,\|\mathbf{h}-\tilde{\mathbf{h}}_{2S}\|_1+\|\mathbf{z}\|_2\right\} \tag{5}$$

*for a given parameter* $\varepsilon$ *. And* $\tilde{\mathbf{h}}_{2S}$ *is a best 2S-sparse approximation to* $\mathbf{h}$.

Following theorem 1, the previously mentioned OMP channel estimation algorithm selects the maximum tap of a sparse channel during one iteration and the channel estimation is carried out in an iterative way. While the proposed CoSaMP SMPC will select the entire dominant taps in each iteration and reduce the estimation error iteration by iteration. Based on the channel model in (1), the CoSaMP SMPC method can be carried out in five steps, as shown in Fig. 2.

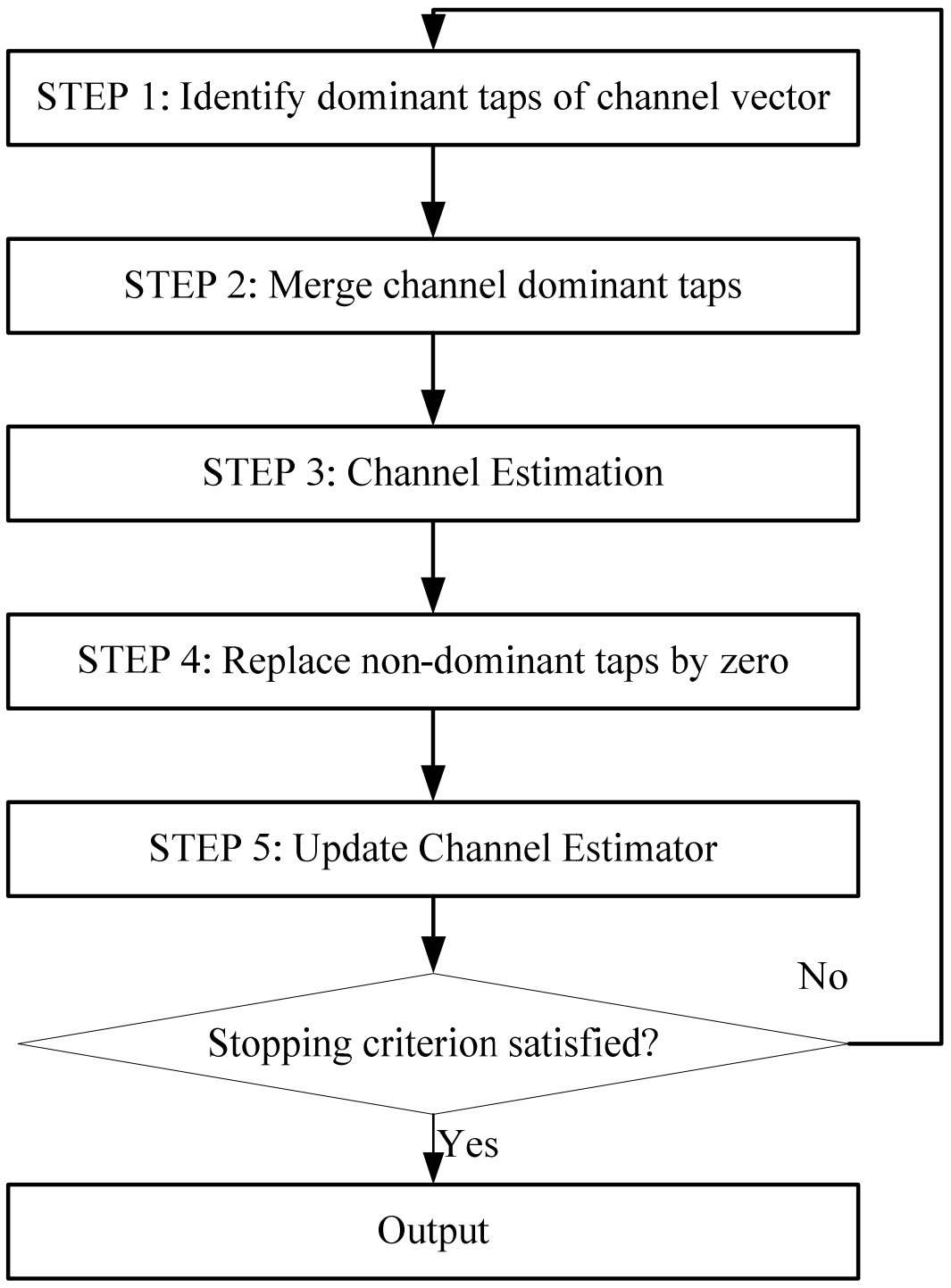

*Fig.2. Steps of CoSaMP SMPC estimation.*

The function of step 1 is to identify the positions of the dominant taps. This step can be divided into two sub-steps. At first, set $P_i = \mathbf{X}^* \mathbf{r}_i$ and choose $2S$ maximum dominant taps. The positions of selected dominant taps are denoted by $\Omega_P$. In the next, using least square (LS) method to calculate a channel estimator as $\tilde{\mathbf{h}}_{LS} = \arg\min \left\| \mathbf{y} - \mathbf{X}\tilde{\mathbf{h}}_{LS} \right\|_2$, and select $S$ maximum dominant taps from $\tilde{\mathbf{h}}_{LS}$. The positions of selected dominant taps in this sub-step are denoted by $\Omega_{LS}$.

The positions of dominant taps are merged by $\Omega_i = \Omega_P \cup \Omega_{LS}$ in step 2.

In step 3, calculate LS channel estimate $\tilde{\mathbf{h}}_i \big|_{\Omega_i} = \mathbf{X}_{\Omega_i}^{\dagger} \mathbf{y}$ on positions in $\Omega_i$.

In step 4, replace the non-dominant taps by zero.

In step 5, update channel estimation error, if the stopping criterion $\left\{ i : i \geq 4S \;\middle|\; \left\| \mathbf{h}_i - \mathbf{h} \right\|_2^2 \leq 10^{-4} \right\}$ is satisfied, then output the channel estimation result; otherwise repeat steps 1~5.

## IV. Simulation Results and Discussion

In this section, the mean square error (MSE) performance of the CoSaMP SMPC estimation method will be evaluated by simulations. For the purpose of comparison, the MSE performance of other existing algorithms such as LS, DS, and OMP algorithms will also be evaluated. In addition, the MSE with LS channel estimation (known position of dominant taps) is also evaluated as a reference.

The simulation condition is listed in Table 1.

*Tab. 1 Simulation condition*

| | | |
|---|---|---|
| Estimation methods | Linear method | LS |
| | CS theory based method | DS |
| | | OMP |
| | | CoSaMP (introduced) |
| Channel fading | Frequency-selective fading | |
| Channel length $L$ | 50 | |
| Taps amplitude | $[-1.0, -0.2] \cup [0.2, 1.0]$ | |
| No. of dominant taps | 5 | |
| SNR | 10dB | |
| Training sequence $\mathbf{X}$ | Toeplitz structure | |
| Length of $\mathbf{X}$ | 15~45 | |

In the following, we will show the results of MSE performance and computational complexity performance according to the CPU time of laptop.

### A. Estimation Error

The MSE is defined as

$$\text{MSE} = \frac{1}{M} \sum_{m=1}^{M} \left\| \mathbf{h} - \tilde{\mathbf{h}}_m \right\|_2^2 \tag{6}$$

It is obvious that smaller MSE means more accurate channel estimation and vice versa. The MSE performance comparisons between the CoSaMP SPMC estimation method and the existing estimation methods are shown in Fig. 3 ~ 4.

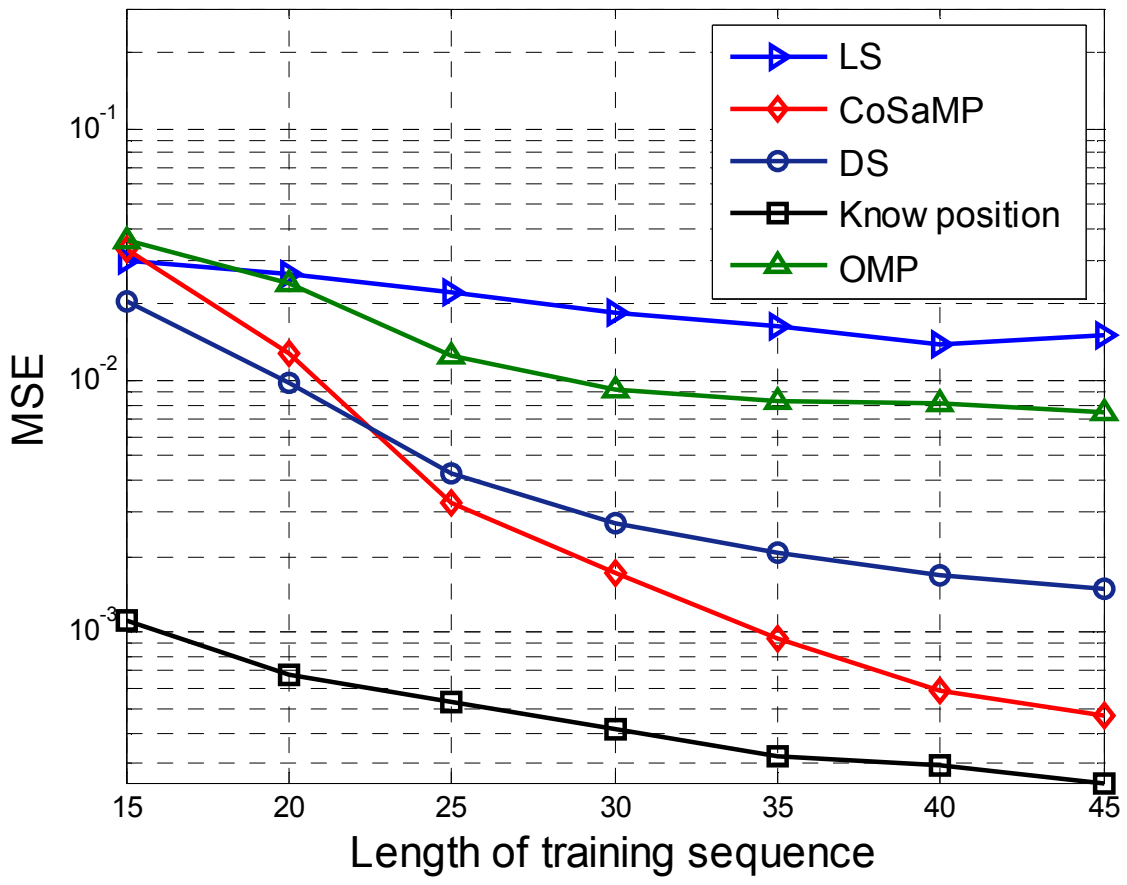

*Fig.3. MSE of the overall taps at SNR=10dB.*

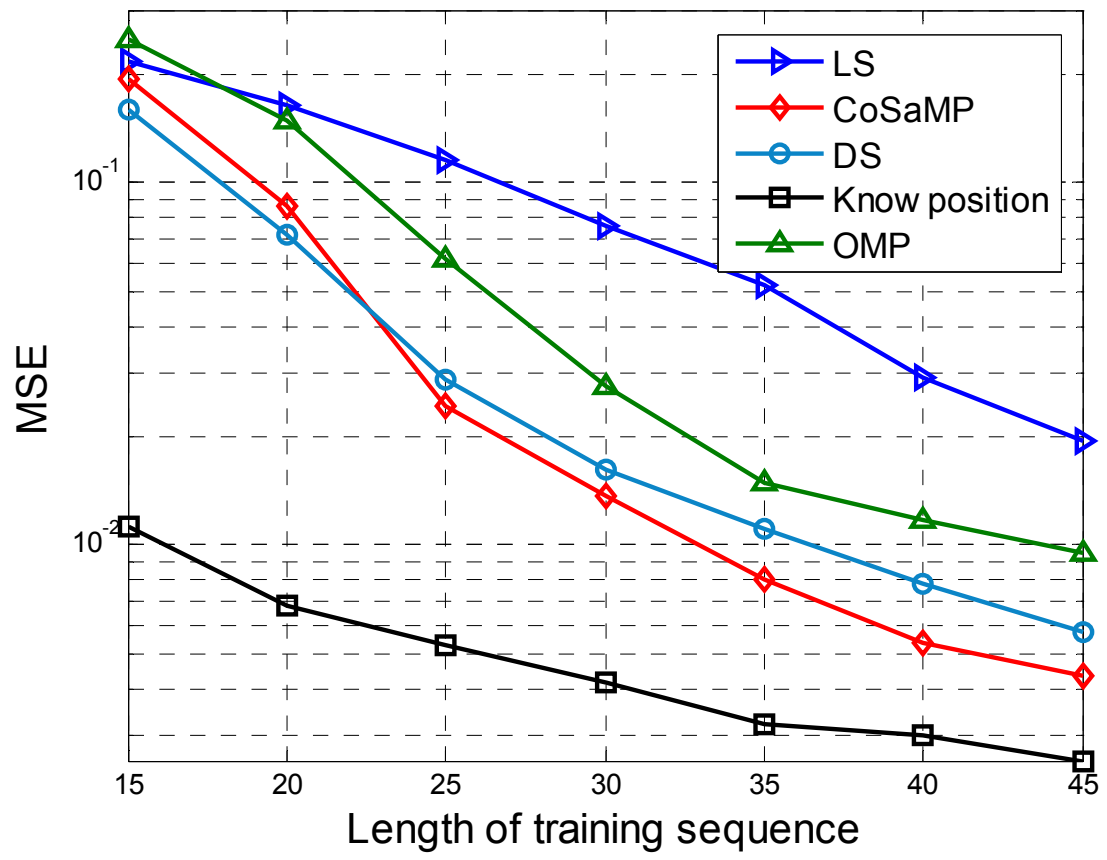

*Fig.4. MSE of the dominant taps at SNR=10dB.*

Fig. 3 shows the result of (6) when all the taps of the channel are considered and Fig. 4 shows the comparison result of (6) when only the dominant taps are considered. It is found that when the training sequence length longer than 25, the CoSaMP method achieves better MSE performance than the other existing methods. This means that the CoSaMP method can achieve the same MSE performance by using shorter training sequence. In other words, the CoSaMP method is more bandwidth efficient.

Here, it was necessary to state that if the length of training sequence less than 25, the MSE performance of CoSaMP is worse than DS which was caused by following reason: DS is a convex optimization algorithm and thus it converts to linear program to resolve. While the CoSaMP is a support set estimation which applies hard threshold by selecting the *S* largest dominant channel taps of a channel vector by applying Least Square (LS) on every iterative step. If the training sequence is very short and then estimate channel unstable on the range of hard threshold. To avoid this deteriorative MSE performance on practical, there have two potential schemes to mitigate. For one thing, we can relax the hard threshold at the cost of acceptable computational complexity. For other thing, we should guarantee lower bound of training sequence length so that robust estimation of the CoSaMP. On the future work, adaptive threshold will consider and further improve channel estimator.

To further study the MSE performance, the cumulative density function (CDF) by using different methods are compared and shown in Fig. 5 ~ 6. It can be clearly observed that the CDF curve of the CoSaMP method is very close to the lower bound (LS channel estimator with known position of dominant taps) and much better than the other estimation methods.

### *B. Rough Estimation of Computational Complexity*

To study the computational complexity of the introduced algorithm, we have evaluated the CPU time in second to complete the channel estimation for $SNR$=10$dB$. It is worth mentioning that although the CPU time is not an exact measure of complexity, it can give us a rough estimation of computational complexity. Our simulations are performance in MATLAB 2007 environment using a 2.40GHz Intel Core-2 processor with 2GB of memory and under Microsoft XP 2003 operating system.

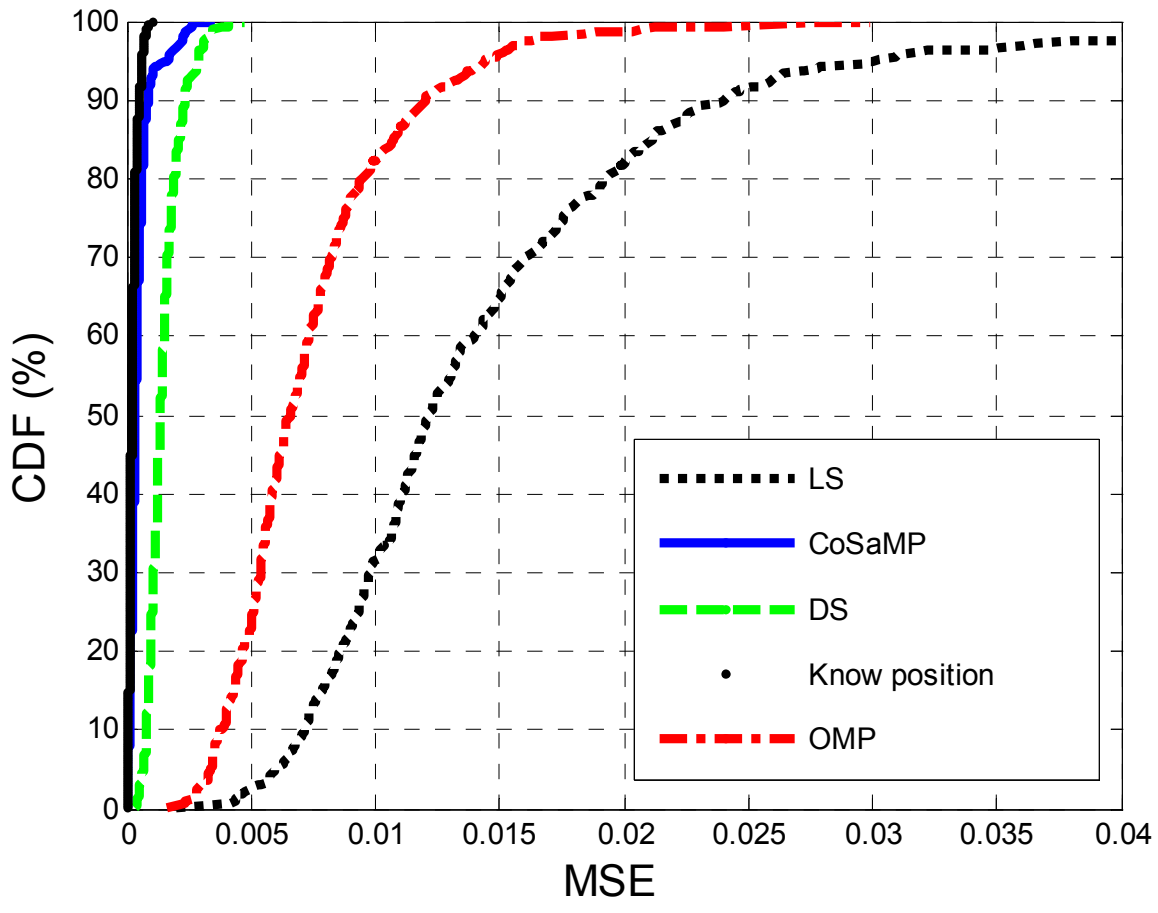

*Fig.5. MSE CDF of the overall taps at SNR=10dB.*

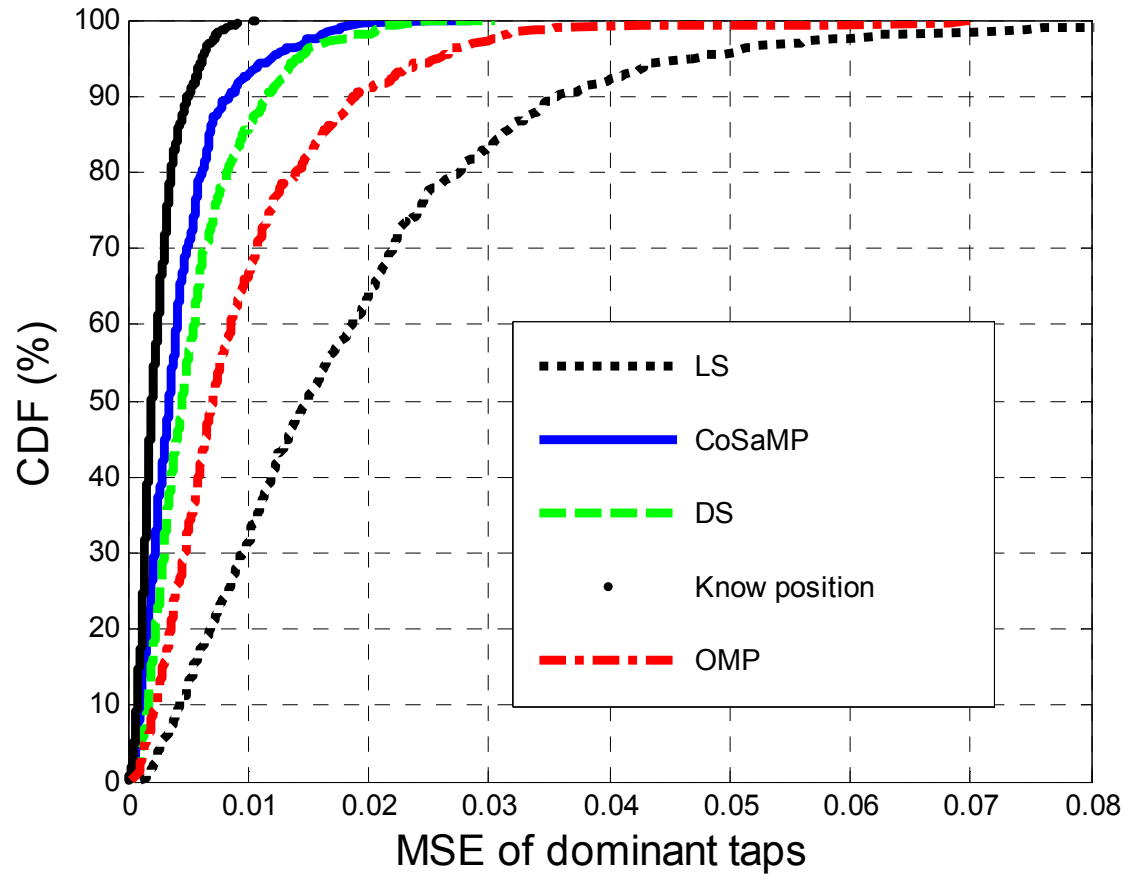

*Fig.6. MSE CDF of the dominant taps at SNR=10dB.*

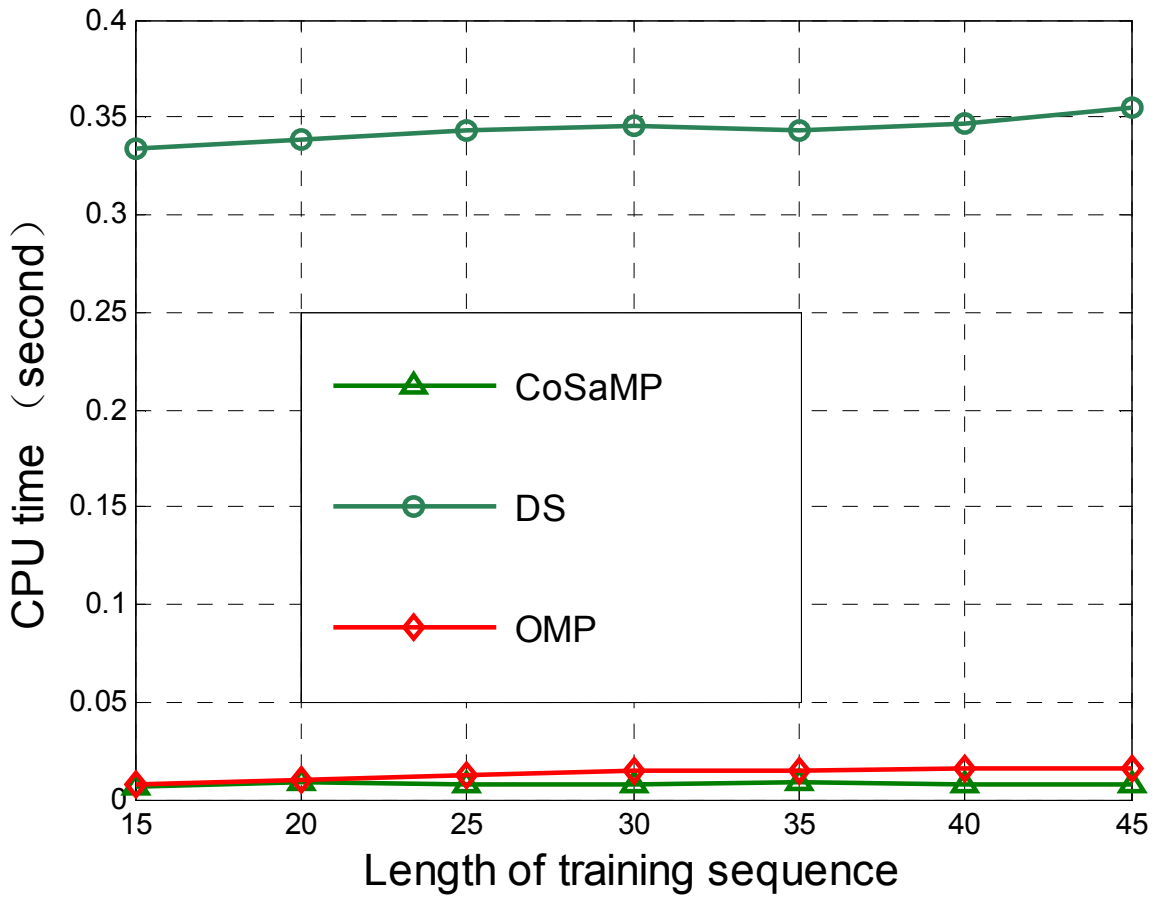

*Fig.7. CPU time for channel estimation*

The comparison between CoSaMP, OMP, and DS algorithms is shown in Fig. 7. It is seen that the computing time of the algorithm is less than 0.05 seconds for both CoSaMP and OMP, while the computing time of the DS algorithm is more than 0.3 seconds. It is shown that the CoSaMP method is also computationally efficient. From the Fig. 7, we also find that CoSaMP take fewer computational time than OMP. Because of the CoSaMP SMPC estimation selects all dominant channel taps while the OMP SMPC estimation chooses maximum channel tap on each iteration.

## V. CONCLUSION AND FUTURE WORK

In this paper, we have introduced a novel sparse channel estimation method based on the CS theory. The CoSaMP method has both advantages of the greedy algorithm and convex program algorithm. It has been shown that, when compared with the existing algorithms, our introduced method is both bandwidth and computationally efficient. However, CoSaMP channel estimation still exist a potential improvement gap. Because of the CoSaMP algorithm considered the hard threshold to choose the set of dominant taps in Fig.2 STEP-2. On future work, we will consider an adaptive CoSaMP channel estimation method which senses the random noise and other unexpected interferences.

## ACKNOWLEDGEMENT

This work is supported in part by the National Natural Science Foundation of China under grant 60772146, the National High Technology Research, Development Program of China (863 Program) under grant 2008AA12Z306 as well as in part by the Key Project of Chinese Ministry of Education under grant 109139, China Scholarship of China Scholarship Council （CSC） under grant No. 2009607029. It is also supported in part by Tohoku University Global COE program "Global Education and Research Center for Earth and Planetary Dynamics".

## REFERENCE

[1] D. Needell and J. A. Tropp, "CoSaMP: Iterative signal recovery from incomplete and inaccurate samples," *Applied and Computational Harmonic Analysis*, vol. 26(3), pp. 301-321, 2008.

[2] Z. Yan, M. Herdin, A. M. Sayeed, and E. Bonek, "Experimental study of MIMO channel statistics and capacity via the virtual channel representation," *Tech. Rep., Univ. Winsconsin-Madison, available http://dune.ece.wisc.edu/pdfs/zhou meas.pdf.*, Feb. 2007.

[3] J. Kivinen, P. Suvikunnas, L. Vuokko, and P. Vainikainen, "Experimental investigations of MIMO propagation channels," *Antennas and Propagation Society International Symposium, IEEE*, 2002.

[4] W. F. Schreiber, "Advanced television systems for terrestrial broadcasting: Some problems and some proposed solutions," *IEEE Proc.*, vol. 83, pp. 958-981, Jun. 1995.

[5] R. Steele, "Mobile Radio Communications," *IEEE Press*, 1992.

[6] M. Kocic, D. Brady, and M. Stojanovic, "Sparse equalization for real time digital underwater acoustic communications," *OCEANS'95, San Diego CA*, pp. 1417-1422, Oct. 1995.

[7] C. Carbonelli, S. Vedantam, and U. Mitra, "Sparse channel estimation with zero tap detection," *IEEE Transactions on Wireless Communnications*, vol. 6(5), pp. 1743-1753, May 2007.

[8] Z. G. Karabulut and A. Yongacoglu, "Sparse channel estimation using orthogonal matching pursuit algorithm," *2004 IEEE 60th Vehicular Technology Conference*, vol. 60(6), pp. 3880-3884, 2004.

[9] J. A. Tropp and A. C. Gilbert, "Signal recovery from random measurements via orthogonal matching pursuit," *IEEE Transaction on Information Theory*, vol. 53(12), pp. 4655-4666, 2007.

[10] U. W. Bajwa, J. Haupt, G. Raz, and R. Nowak, "Compressed channel sensing," *To appear in Proc. 42nd Annu. Conf. Information Sciences and Systems (CISS'08)*, Mar. 19-21, 2008.

[11] D. L. Donoho, M. Elad, and V. N. Temlyakov, "Stable recovery of sparse overcomplete representations in the presence of noise," *IEEE Transaction on Information Theory*, vol. 52(1), pp. 6-18, Jan. 2006.

[12] E. J. Candès, "The restricted isometry property and its implications for compressed sensing," *Compte Rendus de l'Academie des Sciences, Paris*, vol. Serie I, 346, pp. 589-592, 2008.

[13] S. Kunis and H. Rauhut, "Random sampling of sparse trigonometric polynomials II - orthogonal matching pursuit versus basis pursuit," *Found. Comput. Math.*, vol. 8, pp. 737-763, Dec. 2008.

[14] E. Candès and T. Tao, "The Dantzig selector: Statistical estimation when p is much larger than n," *Annals of Statistics*, vol. 35, pp. 2392-2404, 2006.

[15] N. H. Nguyen and T. D. Tran, "The stability of regularized orthogonal matching pursuit algorithm," *http://www.dsp.ece.rice.edu/cs/Stability_of_ROMP.pdf.*

[16] E. Candès, J. Romberg, and T. Tao, "Robust uncertainty principles: Exact signal reconstruction from highly incomplete frequency information," *IEEE Transaction on Information Theory*, vol. 52(2), pp. 489-509, Feb. 2006.